\documentclass{aa}
\input psfig
\begin{document}

   \thesaurus{08       
              (08.16.2;  
               08.09.2;  
               08.18.1;  
               03.20.7)  
               }
   \title{Precise radial velocities of Proxima Centauri.
   \thanks{Based on observations collected at the 
           European Southern Observatory, La~Silla}}
   \subtitle{Strong constraints on a substellar companion}
   \author{M. K\"urster \inst{1} \and
           A. P. Hatzes \inst{2} \and
           W.D. Cochran \inst{2} \and
           S. D\"obereiner \inst{3} \and
           K. Dennerl \inst{3} \and
           M. Endl \inst{1,} \inst{4}
          }
 
   \offprints{M.~K\"urster: mkurster@eso.org}

   \institute{European Southern Observatory, Casilla 19001, Vitacura,
              Santiago 19, Chile
         \and McDonald Observatory, The University of Texas at Austin,
              Austin, TX 78712-1083, USA
         \and Max--Planck--Institut f\"ur extraterrestrische Physik, 
              Giessenbachstr., D-85748 Garching, Germany
         \and Institut f\"ur Astronomie, Universit\"at Wien, 
              T\"urkenschanzstr.~17, A-1180 Wien, Austria}
 
   \date{Received January 20, 1999; accepted February 18, 1999}

   \titlerunning{Precise radial velocities of Proxima Centauri}
   \authorrunning{M. K\"urster et al.}
   \maketitle 

   \begin{abstract}
We present differential radial velocity measurements of Proxima Centauri
collected over 4 years with the ESO CES with a mean precision of 
$54~{\rm ms}^{-1}$. We find no evidence of a periodic signal 
that could corroborate the existence of a sub-stellar companion. 
We put upper limits 
(97\% confidence) to the companion mass ranging from
$1.1$ to $22~{\rm M}_{\rm Jup}$ at orbital periods of $0.75$ to $3000$~d,
i.e.~separations $0.008-2$~AU from Prox~Cen.
Our mass limits concur with limits found by precise astrometry 
(Benedict et~al.~1998a and priv.~comm.) which strongly constrain the period 
range $50-600$~d to $1.1-0.22~{\rm M}_{\rm Jup}$. Combining both results
we exclude a brown dwarf or supermassive planet
at separations $0.008-0.69$~AU from Prox~Cen.
We also find that, at the level of our precision,
the RV data are not affected by stellar activity.
       \keywords{stars: planetary systems --- stars: individual: 
                 Prox~Cen --- stars: rotation ---
                 techniques: radial velocities            }
%
   \end{abstract}
%
\section{Introduction}
Searches for companion objects to Proxima Centauri (Prox~Cen, 
GJ551, V645~Cen, HIP70890), the nearest star ($d=1.2948\pm0.041$~pc; M5Ve), 
date back 20 years to the astrometric work by Kamper \& Wesselink (1978)
and the infrared photometric scanning by Jameson et~al.~(1983).
The most extensive search so far consists 
of astrometric monitoring with the HST FGS \#3 
(Benedict et~al.~1998a) where an astrometric precision of
$0.002^{\prime \prime }$ per axis and a
detection limit for an astrometric variation of 
$0.001^{\prime \prime }$ was achieved,
thereby strongly constraining the mass of a possible companion.
Limits to the $K$-band magnitude of objects within projected separations of 
$1-10$~AU from Prox~Cen were found by Leinert et al. (1997) to be 
$K=12.9-15.1$~mag, i.e.~$\Delta K=3-5$~mag below the empirical end of the 
main-sequence.

Recently the efforts to detect substellar objects near Prox~Cen
culminated in the announcement of a possible
companion by Schultz et~al.~(1998) who used the HST FOS as a
coronographic camera. These authors reported excess light near
Prox~Cen seen in two images separated by 103~d. Within this time span, 
the suspected object appeared to have moved in separation
from $0.23^{\prime \prime}$ to $0.34^{\prime \prime}$ (indicating a separation
near $0.5$~AU) and in P.A.~from $45^{\circ }$ to $100^{\circ }$.
If interpreted as a companion the object would be $\approx 7$~mag fainter 
in the FOS red detector. 

However, a subsequent observation with the HST WFPC2 at two epochs (separated
by 21~d) by Golimowski \& Schroeder (1998) could not verify the existence 
of any companion object to Prox~Cen within a separation from 
$0.09^{\prime \prime}$ to $0.85^{\prime \prime}$ ($0.11-1.1$~AU).
The authors concluded that they should have seen the object at only  
$\approx 3.7$~mag fainter than Prox~Cen in their images taken at $1\mu $m.
Consequently, they suggested that the excess light seen by
Schultz et~al.~(1998) was an instrumental effect.

In this paper we report on 4 years of precise radial velocity (RV) monitoring
of Prox~Cen and contribute new evidence to the debate on a substellar
companion.

\section{The planet search program at the ESO CES}
Our planet search program of 39 late-type stars 
using high-precision RVs was begun at ESO La~Silla in
Nov.~1992. Prox~Cen was first observed in July 1993. 
We used the ESO 1.4m CAT 
telescope and CES spectrograph equipped with the f/4.7 Long Camera 
and ESO CCDs $\# 30$ or $\# 34$. The obtained resolving power, central
wavelength and spectrum length were $100,000$, $5389$~{\AA }, and $48$~{\AA }.
For high measurement precision for differential RVs we self-calibrated the 
spectrograph using an iodine ($I_2$) gas absorption cell temperature 
controlled at $50^{\circ }$~C (K\"urster et~al.~1994; Hatzes et~al.~1996; 
Hatzes \& K\"urster 1994).
To obtain RV measurements we model the stellar
spectra as observed through the iodine cell using a `pure'
stellar spectrum (recorded without the iodine cell in the light path) 
and a `pure' iodine ($I_2$) spectrum from dome flat measurements.
The resulting RV data are then corrected to the solar system barycenter
via the JPL ephemeris DE200.

For stars brighter than 5.5~mag our short-term (i.e.~single night), best case 
long-term, and working long-term precisions (i.e.~obtained under all observing
conditions) are $4-7$, $11$, and $20-25~{\rm ms}^{-1}$, respectively 
(K\"urster et~al.~1994, 1998, 1999). Prox~Cen is unique in our sample
in that it is by  far the faintest star we have observed ($V=11.01$~mag). 

%
\section{The RV measurements for Proxima Cen}
Tab.~1 shows the journal of observations of our differential RV measurements.
A total of 58 spectra from 29 nights
were available. Before further analysis the spectra were combined into
nightly bins (col.~1) as outlined below,
with the bins containing between 1 and 5 spectra (col.~2).

While self-calibration with an iodine cell is an excellent method to overcome
instrumental instabilities such as instrumental drifts other
instabilities such as focus and alignment changes or instrument vibrations
require (in principle) additional modelling of the instrumental
profile (IP; Butler et~al.~1996; Valenti et~al.~1995).
At low signal-to-noise (S/N) ratios such as obtained
for Prox~Cen IP reconstruction becomes impossible.
However, to some extent one can overcome the remaining
measurement uncertainty by modelling the observed spectra (star+iodine)
with various combinations of pure star and iodine spectra.

Ten different pure star and pure iodine pairs (of the same night) were
available to build models for each of the 58 star+iodine spectra. Based on
goodness-of-fit some of the most inadequate models were rejected. 
%
Col.~3 gives the total number of the accepted star+iodine models for all
the spectra in each nightly bin (sum over all spectra and models).
%
For an intercomparison of the different models their RV zero points have to be 
matched. To do this we subtracted for each model the mean RV 
for the whole times series. Subsequently, we averaged for each star+iodine
spectrum the RV data from the individual models.
At last, nightly (bin) averages for the Julian day (col.~4) and  RV 
data corrected to zero mean (col.~5) were calculated. 

A first estimate of the RV error (col.~6) was then based on 
the rms scatter in each data bin together with a propagation of the error
introduced by the process of matching the zero points for the different models.
%
However, we found a positive correlation with a correlation coefficient of 
$+0.514$ between the total number of models in a bin (col.~3) and the error
estimate (col.~6) meaning that these errors tend to be smaller when
estimated from smaller numbers of models. This indicates that these error 
estimates are not representative of the true errors,
in particular those from small numbers of models.
%

As an independent estimate of the weight of each data bin, col.~7 shows the 
combined S/N ratio per spectral pixel (mean values) of each data bin.
We know from simulations (Hatzes \& Cochran 1992) that the RV measurement
error $\Delta RV\propto (S/N)^{-1}$
which serves to estimate the RV error.
Choosing the constant of proportionality such that the mean of the resulting
errors is equal to the total scatter in the RV measurements 
($53.9~{\rm ms}^{-1}$) we obtain
the equivalent RV errors listed in col.~8 of Tab.~1.
Fig.~1 displays a time series of our RV data (29 bins)
for Prox~Cen with error bars corresponding to these equivalent errors. 

%
   \begin{table}
      \caption[]{
%
RV data for Prox~Cen. Col.~1: bin number; col.~2: number of
spectra in the bin; col.~3: total number of models for all spectra in  
the bin (i.e.~the sum over all spectra and models); col.~4: mean 
Julian day of the bin; col.~5: mean differential RV of the bin;  
col.~6: rms scatter within the bin; col.~7: combined S/N ratio of the
bin; col.~8: equivalent RV error as estimated from the S/N.
%
         }
         \begin{flushleft}
            \begin{tabular}{ccrccc}
\hline
\noalign{\smallskip}
\#  \#  $\Sigma $ & JD $-$ & $\Delta $RV & rms & comb & err$_{\rm eq}$ \\
Sp            & $2,400,000$ 
 & $[$m/s$]$ & $[$m/s$]$ & S/N & $[$m/s$]$ \\
\noalign{\smallskip}
\hline
\noalign{\smallskip}
~1  1  ~7 & 49179.6643 & $~106.35$ & $~44.85$ & 21.46 & 71.78 \\
~2  2  14 & 49246.5034 & $~-18.00$ & $~78.30$ & 17.88 & 86.16 \\
~3  2  18 & 49412.8484 & $~~~3.48$ & $~46.24$ & 23.87 & 64.54 \\
~4  4  34 & 49413.7676 & $~-78.83$ & $~68.73$ & 34.10 & 45.18 \\
~5  5  40 & 49492.7022 & $~~52.55$ & $~85.96$ & 39.78 & 38.73 \\
~6  4  30 & 49548.5800 & $~~56.66$ & $~65.18$ & 35.12 & 43.86 \\
~7  2  15 & 49602.5149 & $~~-2.23$ & $~60.60$ & 28.44 & 54.17 \\
~8  2  17 & 49794.7186 & $~107.13$ & $103.27$ & 26.28 & 58.62 \\
~9  2  14 & 49795.6955 & $~~28.33$ & $~44.09$ & 28.10 & 54.82 \\
10  2  16 & 49906.5524 & $~~~0.64$ & $~44.11$ & 27.91 & 55.20 \\
11  1  ~9 & 49907.5940 & $~~60.47$ & $~38.37$ & 21.47 & 71.75 \\
12  1  ~8 & 50146.7675 & $~-70.90$ & $~42.87$ & 25.89 & 59.50 \\
13  2  17 & 50229.6486 & $~-40.73$ & $~56.75$ & 36.54 & 42.16 \\
14  1  ~8 & 50230.6858 & $~~55.73$ & $~49.09$ & 26.82 & 57.44 \\
15  2  14 & 50304.4963 & $~~~2.19$ & $~52.60$ & 30.92 & 49.82 \\
16  2  15 & 50313.5397 & $~-35.71$ & $~43.80$ & 40.39 & 38.14 \\
17  2  14 & 50314.5965 & $~-37.08$ & $~49.38$ & 37.62 & 40.95 \\
18  2  13 & 50319.5519 & $~-21.17$ & $~45.65$ & 31.42 & 49.03 \\
19  1  ~6 & 50323.5606 & $~-88.32$ & $~30.21$ & 25.87 & 59.55 \\
20  1  ~7 & 50325.5852 & $~-47.33$ & $~36.42$ & 27.10 & 56.85 \\
21  2  17 & 50348.5179 & $~~~6.32$ & $~45.15$ & 38.82 & 39.68 \\
22  2  18 & 50349.5142 & $~-18.19$ & $~49.07$ & 36.90 & 41.75 \\
23  1  ~7 & 50359.4920 & $~-36.71$ & $~42.60$ & 24.56 & 62.72 \\
24  2  17 & 50477.8459 & $~~45.31$ & $~73.70$ & 25.24 & 61.03 \\
25  2  14 & 50478.8507 & $-104.59$ & $~93.69$ & 21.67 & 71.09 \\
26  2  18 & 50524.7935 & $~~43.71$ & $~48.86$ & 28.80 & 53.49 \\
27  2  17 & 50551.6664 & $~~44.23$ & $~48.27$ & 28.03 & 54.96 \\
28  2  12 & 50570.7647 & $~~-3.27$ & $~61.04$ & 38.30 & 40.22 \\
29  2  15 & 50648.5032 & $~-10.02$ & $~42.17$ & 38.09 & 40.44 \\
\noalign{\smallskip}
\hline
\end{tabular}
\vskip -0.5cm
\end{flushleft}
\end{table}
%

%
\section{Period search}
To look for a periodic signal that could manifest the presence of a companion,
we searched the frequency range 
$f_{\rm min}=1/T~...~f_{\rm max}=1/(2\Delta t)$,
or $0.0007~...~0.5494~{\rm d}^{-1}$,
where $T$ is the total time baseline 
and $\Delta t$ is the minimum separation between data points. Thus a period
range of $1.8202~...~1428.6$~d was searched. The 
choice of the maximum frequency was made in analogy to
the Nyquist criterion which, however, is well defined only for equidistant
data sampling. 
Since most of our data sampling is much cruder than our minimum
sampling any signals at periods shorter than about 5~d should be treated 
with care.

%
\begin{figure}
\centering{
  \vbox{\psfig{figure=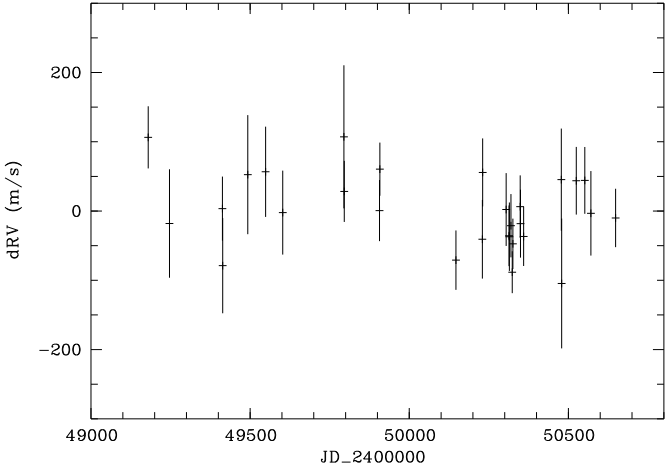,width=8.8cm,%
      bbllx=1.5cm,bblly=1.0cm,bburx=17.8cm,bbury=12.7cm,clip=}}\par
      }
  \caption{Time series of the differential RV data of Prox~Cen.
            }
\end{figure}
%

Two different types of periodogram were used: a) the Scargle periodogram
(Scargle 1982) which is equivalent to least squares sine fitting with
equal weight for all data points; the power
is proportional to the square of the amplitude of the corresponding sine wave;
b) a sine-fitting routine that takes into account data errors by minimizing
$\chi ^2$; we used the equivalent errors (col.~8 of Tab.~1).

Fig.~2 shows only the Scargle periodogram, since the $\chi ^2$ minimization
approach yielded a very similar result that does not change the interpretation.
>From 10,000 runs of a bootstrap randomization scheme (see 
K\"urster et~al.~1996; Murdoch et~al.~1993) we determined the levels of 
the false alarm probability $\phi $ (FAP) corresponding to various power
levels. As shown in Fig.~2 all periodogram peaks are insignificant having 
$\phi >90\% $.

In contrast to searching a {\em period range} signals {\em at a priori known 
periods} can be significant at smaller power levels $z$.
For the Scargle periodogram the FAP is given by $\phi =1-(1-e^{-z})^n$,
where $n$ is the number of independent frequencies in the 
search interval (Scargle 1982).
Hence for a signal at a single a priori known period $\phi =e^{-z}$.

Periods that may be present are the period of the stellar rotation and that
of the
activity cycle. RV searches strongly benefit from ancilliary 
information such as the knowledge of these periods that aids 
the interpretation of RV data.
Star spots and inhomogeneous granulation patterns in active stars cause
distortions in stellar absorption lines; when
sufficient resolution and/or signal-to-noise is lacking these distortions can 
be mis-interpreted as RV shifts that vary with the rotation period 
(rotational modulation) or with the activiy cycle.
Estimates for the rotation period of Prox~Cen were given by Benedict 
et~al.~(1998b; based on HST FGS photometry) finding
$P_{\rm rot}=83.5$~d plus variability at the first harmonic,
and by (Guinan \& Morgan 1996; monitoring of the MgII~h+k flux with IUE)
who found $31.5\pm 1.5$~d. Benedict et~al.~(1998b)
also estimate the length of Prox~Cen's activity cycle to be $\approx 1100$~d.

%
\begin{figure}
\centering{
  \vbox{\psfig{figure=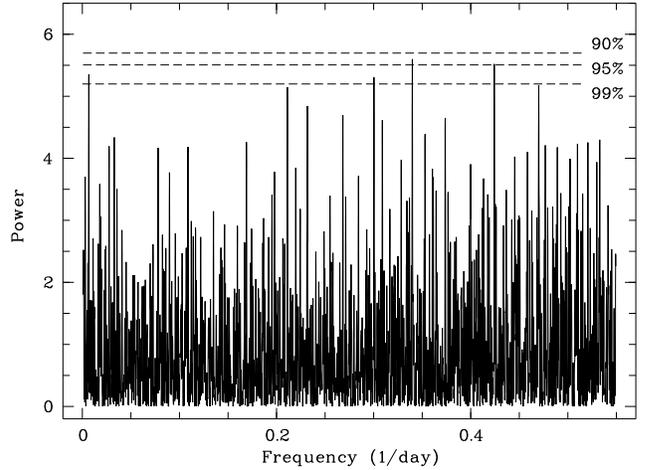,width=8.8cm,%
      bbllx=1.5cm,bblly=1.0cm,bburx=17.8cm,bbury=12.7cm,clip=}}\par
      }
  \caption{Periodogram for the RV data.
           Horizontal dashed lines correspond to various levels of the false 
           alarm probability.} 
\end{figure}
%

We do not find significant FAPs at any one of these periods
nor at first harmonics or periods twice as large
(relevant in case the literature values are first harmonics themselves).
Allowing for some uncertainty in these period values
we also searched their vicinities finding 
$\phi =0.47\%~(0.16\%)$ for $P=150.4~d$ and
$\phi =1.32\%~(0.92\%)$ for $P=30.4~d$, where theoretical values,
$\phi =e^{-z}$, as well as
values bootstrapped from 10,000 runs (values in brackets) are given.
At best, i.e.~only if one allows for considerable errors in the original period
estimates, a marginal detection of
(1) a period twice as long as the rotation period by 
Benedict et~al.~(1998b) and (2) the rotation period
by (Guinan \& Morgan 1996) might be indicated.

Reconfiguration of active regions causes amplitude and
phase changes in rotationally modulated signals complicating their detection
in data that extend over many rotation cycles.
However, our results concur with the findings by Saar et~al.~(1998)
that predict an activity-related RV scatter of  
$<10~{\rm ms}^{-1}$ for rotation periods $>16$~d.
It appears that the RV variation seen for Prox~Cen is 
representative of our measurement precision for this faint star, and
cannot be attributed to instrinsic stellar variability.

\section{Limits to companion parameters}

Lacking a clear RV signal in our Prox~Cen data we used a Monte
Carlo simulation to derive upper limits to the mass of still possible
companions in the period range $0.75-3000$~d.
Random data sets with the same temporal sampling and with the rms scatter 
of our data were created.
We added sinusoidal signals with different periods,
amplitudes, and phases, and evaluated their periodograms. At each
period the amplitude was determined for which 99\% of the 
periodograms showed a power corresponding to $\ge 1\% $ FAP.
Hence the combined confidence for the detection of a sinusoidal 
signal is 98\%. For eccentric orbits it may be somewhat lower.

%
\begin{figure}
\centering{
  \vbox{\psfig{figure=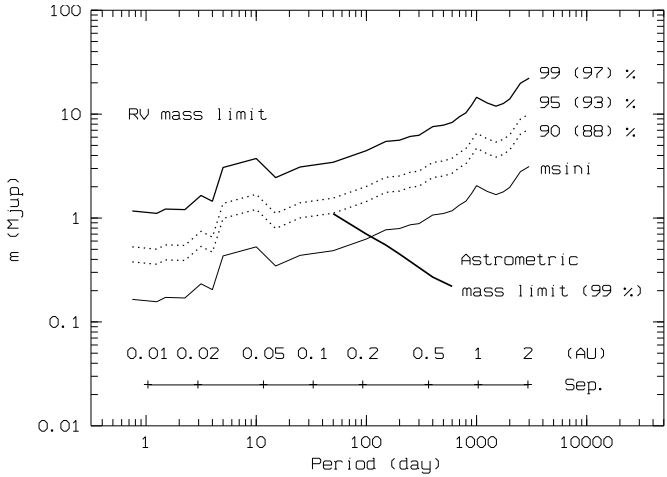,width=8.8cm,%
      bbllx=1.5cm,bblly=1.0cm,bburx=17.8cm,bbury=12.7cm,clip=}}\par
      }
  \caption{RV derived upper limits to the mass of companions
           to Prox~Cen as a function of orbital period and separation. The 
           four parallel curves represent the $m\sin i$ 
           and several confidence intervals for the mass limit.
           The single bold line is the mass upper limit 
           from HST FGS astrometry (Benedict, priv.~comm.).}
\end{figure}
%

Fig.~3 shows the $m\sin i$ ($m$ the planet mass, $i$ the orbital inclination),
that corresponds to RV amplitudes at this confidence level, as a function of 
period and separation. We assumed a stellar mass of $0.12~{\rm M}_\odot $
(for an M5Ve star; Kirkpatrick \& McCarthy 1994).
To account for the unknown inclination one can use its
probability distribution (for random orientation
of the orbits). The probability that $i$ exceeds some 
angle $\theta $ is given by $p(i>\theta ) = \cos(\theta )$. From this one can
construct confidence intervals for the true companion mass. There is a
confidence of 90\% (95\%, 99\%) that the true mass is no more than a
factor 2.294 (3.203, 7.088) larger than the $m\sin i$. Mass limits for these
confidence intervals are also included in Fig.~3 with the corresponding 
labels. Values in brackets are the combined confidence levels accounting for 
both the confidence of the inclination and the confidence of detection.
We choose the curve with the highest confidence (97\% combined) as the
RV derived upper mass limit.

We also got upper limits (99\% confidence) to the companion mass 
from HST FGS astrometry (Benedict, priv. comm.; 
cf.~Benedict et~al.~1998a) that are included in Fig.~3.
Being most stringent at longer periods they complement our RV derived limits 
constraining the period range $50-600$~d.
When combining them with our RV mass limits
we can exclude massive planets around Prox~Cen over a wide period range
as detailed in Sect.~6.

\section{Conclusions}
\begin{enumerate}
\item Prox~Cen does not have a close ($\approx 0.4$~AU) brown dwarf companion
      as suggested by Schultz et al.~(1998).
\item RV derived upper mass limits range from $1.1$ to 
      $3.7~{\rm M}_{\rm Jup}$ for periods from a few days to a few weeks.
\item In the period range $50-600$~d (separations $0.13-0.69$~AU) 
      the RV derived mass limits range from $3.4$ to $8.3~{\rm M}_{\rm Jup}$;
      in this interval the more stringent astrometry even
      indicates the absence of objects from $1.1$ to $0.22~{\rm M}_{\rm Jup}$, 
      i.e.~below Saturn mass for periods $\ge 370$~d.
\item Hence no massive planets $>3.7~{\rm M}_{\rm Jup}$ exist in
      orbits with periods of $0.75-600$~d, i.e.~at $0.008-0.69$~AU.
\item At periods $>600-3000$~d (separations $>0.69-1$~AU) RV derived 
      mass limits range from $8.3$ to $22~{\rm M}_{\rm Jup}$.
\item At the level of our measurement precision the RV data are not notably
      affected by stellar activity.
\end{enumerate}
%

\begin{acknowledgements}
We are grateful to F.~Benedict for communicating to us the astrometric
mass limits.
We thank the ESO OPC for generous allocation of
observing time. The support of the La Silla 3.6m+CAT team and the
Remote Control Operators at ESO Garching was invaluable for obtaining 
these data. 
APH and WDC acknowledge support by NASA grant NAG5-4384 and
NSF grant AST-9808980.
\end{acknowledgements}


%
\end{document}